\documentclass[epj]{svjour}
\usepackage{graphics}
\usepackage{amssymb}
\newcommand\als{\alpha_{\rm s}}
\newcommand\lQ{\Lambda_{QCD}}
\newcommand{\be}{\begin{equation}}
\newcommand{\ee}{\end{equation}}
\newcommand{\bea}{\begin{eqnarray}}
\newcommand{\eea}{\end{eqnarray}}
\newcommand{\nn}{\nonumber}
\def\bfsigma{\mbox{\boldmath $\sigma$}}

\begin{document}
\title{Overview of Non-Relativistic QCD}
\author{Joan Soto
}                     
%
%
\institute{Departament d'Estructura i Constituents de la Mat\`eria, Universitat de Barcelona}
\date{Received: 15/11/2006}
%
\abstract{
An overview of recent theoretical progress on Non-Relativistic QCD and related effective theories is provided.
      } 
\maketitle
\section{Introduction}
\label{intro}
Heavy quarkonium systems have been a good laboratory to test our theoretical ideas since the early days of QCD (see \cite{Brambilla:2004wf} for an extensive review). Based on the pioneering work of Caswell and Lepage \cite{Caswell:1985ui}, a systematic approach to study such systems from QCD has been developed over the years which makes use  of effective field theory techniques and is generically known as Non-Relativistic QCD (NRQCD) \cite{Bodwin:1994jh}. NRQCD has been applied to spectroscopy, decay and production of heavy quarkonium. The NRQCD formalism for spectroscopy and decay is very well understood (this is also so for electromagnetic threshold production)
so that NRQCD results may be considered QCD results up to a given order in the expansion parameters (usually $\als (m_Q)$ and $1/m_Q$, $m_Q$ being the heavy quark mass).
I will restrict myself to review recent progress on these issues and refer the reader to the review \cite{Brambilla:2004jw} for earlier developments. The NRQCD formalism for production is more controversial. 
Important work has been carried out recently to which I will devote some time. 
\section{Heavy Quarkonium}
\label{hq}
Heavy quarkonia are mesons made out of a heavy quark and a heavy antiquark (not necessarily of the same flavor), whose masses are larger than $\lQ$, the typical hadronic scale. These include bottomonia ($b\bar b$), charmonia ($c \bar c$), $B_c$ systems ($b \bar c$ and $c \bar b$) and would-be toponia ($t\bar t$). Baryons made out of two or three heavy quarks
share some similarities with these systems (see \cite{Brambilla:2005yk,Fleming:2005pd}).

In the quarkonium rest frame the heavy quarks move slowly ($v \ll 1$, $v$ being the typical heavy quark velocity in the center of mass frame), with a typical momentum $m_Qv \ll m_Q$
and binding energy $\sim m_Qv^2$. Hence any study of heavy quarkonium faces a multiscale problem with the hierarchies $m_Q \gg m_Qv \gg m_Qv^2$ and $m_Q \gg \lQ$. The use of effective field theories is extremely convenient in order to exploit these hierarchies.
\section{Effective Field Theories}
\label{efts}

Direct QCD calculations in multiscale problems are extremely difficult, no matter if one uses analytic approaches (i.e. perturbation theory) or numerical ones (i.e. lattice). We may try to construct a simpler theory (the effective field theory (EFT)), in which less scales are involved in the dynamics and which is equivalent to the fundamental theory (QCD) in the particular energy region where heavy quarkonia states lie. The clues for the construction are: (i) identify the relevant degrees of freedom, (ii) enforce the QCD symmetries and (iii) exploit the hierarchy of scales. Typically one integrates out the higher energy scales so that the lagrangian of the EFT can be organized as a series of operators over powers of these scales. Each operator has a so called matching coefficient in front, which encodes the remaining information on the higher energy scales. The matching coefficients may be calculated by imposing that a selected set of observables coincide when are calculated in the fundamental and in the effective theory.  

NRQCD is the (first) effective theory relevant for heavy quarkonium. Out of the four components of the  relativistic Dirac fields describing the heavy quarks (antiquarks) only the upper (lower) are relevant for energies lower than $m_Q$ (no pair production is allowed anymore). Hence a two component Pauli spinor field is used to describe the quark (antiquark). The hierarchy of scales exploited in NRQCD is $m_Q \gg m_Qv , m_Qv^2 , \lQ$. The remaining hierarchy $m_Qv \gg m_Qv^2$, may be exploited using a further effective theory called Potential NRQCD (pNRQCD) \cite{Pineda:1997bj,Brambilla:1999xf}.  

\subsection{Non-Relativistic QCD}
\label{NRQCD}

The part of the NRQCD lagrangian bilinear on the heavy quark fields coincides with the one of Heavy Quark Effective Theory  (HQET) (see \cite{Neubert:1993mb} for a review), and reads
\bea
&&{\cal L}_{\psi}=
\psi^{\dagger} \Biggl\{ i D_0 + {1\over 2 m_Q} {\bf D}^2 + {1 \over 8 m_Q^3} {\bf D}^4 
+ {c_F \over 2 m_Q} {\bfsigma . g{\bf B}} \, 
+ \nn\cr && + { c_D \over 8 m_Q^2} \left({\bf D} . g{\bf E} - g{\bf E} . {\bf D} \right) +
 i \, { c_S \over 8 m_Q^2} 
{\bfsigma . \left({\bf D} \times g{\bf E} -g{\bf E} \times {\bf D}\right) }
\Biggr\} \psi\label{hqet}
\eea
$\psi$ is a Pauli spinor which annihilates heavy quarks. $c_F$, $c_D$ and $c_S$ are short distance matching coefficients which depend on $m_Q$ and $\mu$ (factorization scale). Analogous terms exist for $\chi$, a Pauli spinor field which creates antiquarks. Unlike HQET, it also contains four fermion operators,

\bea 
{\cal L}_{\psi\chi}& =&
  {f_1(^1S_0) \over m_Q^2}O_1(^1S_0) 
+ {f_1(^3S_1) \over m_Q^2}O_1(^3S_1) +\nn\\ && 
+ {f_8(^1S_0) \over m_Q^2} O_8(^1S_0)
+ {f_8(^3S_1) \over m_Q^2} O_8(^3S_1),
\nn
\eea
\bea 
O_1(^1S_0)&=\psi^{\dag} \chi \, \chi^{\dag} \psi &, 
\quad O_1(^3S_1)=\psi^{\dag} {\bfsigma} \chi \, \chi^{\dag} {\bfsigma} \psi , 
\label{def4fops}
\\
O_8(^1S_0)&=\psi^{\dag} {\rm T}^a \chi \, \chi^{\dag} {\rm T}^a \psi &, 
\quad O_8(^3S_1)= \psi^{\dag} {\rm T}^a {\bfsigma} \chi \, \chi^{\dag} {\rm T}^a {\bfsigma} \psi .
\nn
\eea
The $f$s are again short distance matching coefficients which depend on $m_Q$ and $\mu$,
which have imaginary parts when the quark and the antiquark are of the same flavor. This is due to the fact that hard gluons (of energy $\sim m_Q$), which remain accessible through annihilation of the quark and antiquark, have been integrated out. The fact that NRQCD is equivalent to QCD at any desired order in $1/m_Q$ and $\als (m_Q)$ makes the lack of unitarity innocuous. In fact, it is turned into an advantage: it facilitates the calculation of inclusive decay rates to light particles.

The lagrangian above can be used as such for spectroscopy, inclusive decays and electromagnetic threshold production of heavy quarkonia. Spectroscopy studies have been carried out on the lattice and are discussed in \cite{Vairo:2006nq} (see also \cite{Gray:2005ur} and references therein).  

\subsubsection{Inclusive decays} 

Let us just show, as an example, the NRQCD formula for inclusive decays of P-wave states to light hadrons at leading order

\bea
&& \Gamma(\chi_Q(nJS)  \rightarrow LH)= 
{2\over m_Q^2}\Bigg( {\rm Im \,}  f_1(^{2S+1}P_J) \times \nn\\ && \times
{\langle \chi_Q(nJS) | O_1(^{2S+1}P_J ) | \chi_Q(nJS) \rangle \over m_Q^2}
\nn
\\ &&
+ {\rm Im \,} f_8(^{2S+1}S_S) \langle \chi_Q(nJS) | O_8(^1S_0 ) | \chi_Q(nJS) \rangle\Bigg),\nn
\eea
$f_1$ and $f_8$ are short distance matching coefficients which can be calculated in perturbation theory in $\als (m_Q)$. $O_1$ $(^{2S+1}P_J)$ is a color singlet dimension 8 operator and $O_8(^1S_0 )$ is the color octet operator given in (\ref{def4fops}). Earlier QCD factorization formulas 
were missing the color octet contribution. They are inconsistent because the color octet matrix element is necessary to cancel the factorization scale dependence which arises in one loop calculations of ${\rm Im \,}  f_1$ $(^{2S+1}P_J)$ \cite{Bodwin:1992ye}. The matrix elements cannot be calculated in perturbation theory of $\als (m_Q)$. They can however be calculated on the lattice (see for instance \cite{Bodwin:2001mk}) or extracted from data. The imaginary parts of the matching coefficients of the dimension 6 and 8 operators are known at NLO \cite{Petrelli:1997ge}, and those of the electromagnetic decays of dimension 10 at LO \cite{Brambilla:2006ph,Bodwin:2002hg,Ma:2002ev}.

\subsubsection{Problems}

Unlike HQET, in which each term in the lagrangian (\ref{def4fops}) has a definite size (by assigning $\lQ^n$
to any operator of dimension $n$), NRQCD does not enjoy a homogeneous counting. This is due to the fact that the scales $m_Qv$, $m_Qv^2$ and $\lQ$ are still entangled. Even though a counting was put forward in \cite{Bodwin:1994jh}, the so called NRQCD velocity counting, which has become the standard bookkeeping for NRQCD calculations, there is no guarantee that it holds for all the states, and hence there is a problem on how to make a sensible organization of the calculation for a given state.  There are two approaches which aim at disentangling these scales, and hence to facilitate the counting. The first one consist of constructing a further effective theory by integrating out energy scales larger $m_Q v^2$. This is the pNRQCD approach first proposed in \cite{Pineda:1997bj}, which will be described in the next section. The second one consist in decomposing the NRQCD fields in several modes so that each one has a homogeneous counting. This is the vNRQCD approach first proposed in \cite{Luke:1999kz} (see also \cite{Griesshammer:1997wz}). The scale $\lQ$ has not been discussed so far in this approach, and, in fact, a consistent formulation has only become available recently \cite{Manohar:2006nz} (see also \cite{Hoang:2002yy,Manohar}).

\subsection{Potential NRQCD}

As mentioned above, pNRQCD is the effective theory which arises from NRQCD after integrating out energy scales largen than $m_Q v^2$, namely than the typical binding energy. If $\lQ \lesssim m_Q v^2$, then $m_Q v \gg \lQ$ and the matching between NRQCD and pNRQCD can be carried out in perturbation theory in $\als (m_Qv)$ . This is the so called weak coupling regime. If $\lQ \gg m_Q v^2$, the matching cannot be carried out in perturbation theory in $\als (m_Qv)$ anymore, but one can still exploit the hierarchy $m_Q \gg m_Q v , \lQ \gg m_Q v^2$. This is the so called strong coupling regime.

\subsubsection{Weak Coupling Regime}

The lagrangian in this regime reads
\bea
& & \!\!\!\!\!\!\!
{\cal L}_{\rm pNRQCD} = \int d^3{\bf r} \; {\rm Tr} \,  
\Biggl\{ {\rm S}^\dagger \left( i\partial_0 
- h_s({\bf r}, {\bf p}, {\bf P}_{\bf R}, {\bf S}_1,{\bf S}_2,\mu ) \right) {\rm S} +\nn\\ &&\qquad\qquad 
+ {\rm O}^\dagger \left( iD_0 
- h_o({\bf r}, {\bf p}, {\bf P}_{\bf R}, {\bf S}_1,{\bf S}_2,\mu ) \right) {\rm O} \Biggr\}
\nn
\\
& &\qquad\qquad 
+ V_A ( r, \mu ) {\rm Tr} \left\{  {\rm O}^\dagger {\bf r} \cdot g{\bf E} \,{\rm S}
+ {\rm S}^\dagger {\bf r} \cdot g{\bf E} \,{\rm O} \right\} + \nn\\&&\qquad\qquad 
+ {V_B (r,\mu ) \over 2} {\rm Tr} \left\{  {\rm O}^\dagger {\bf r} \cdot g{\bf E} \, {\rm O} 
+ {\rm O}^\dagger {\rm O} {\bf r} \cdot g{\bf E}  \right\} \nn 
\eea
where $S$ and $O$ are color singlet and color octet wave function fields respectively. $h_s$ and $h_o$ are color singlet and color octet hamiltonians respectively, which may be obtained by matching to NRQCD in perturbation theory in $\als (m_Qv)$ and $1/m_Q$ at any order of the multipole expansion (${\bf r}\sim 1/m_Qv$). The static potential terms in $h_s$ and $h_o$ are known at two loops \cite{Peter:1997me,Schroder:1998vy,Kniehl:2004rk}
(the logarithmic contributions at three loops and (for $h_s$ only) at four loops  are also known \cite{Brambilla:1999qa,Brambilla:1999xf,Brambilla:2006wp}). The renormalon singularities in the static potentials are also understood in some detail \cite{Pineda:2002se,Bali:2003jq}. The $1/m_Q$ and $1/m_Q^2$ terms in $h_s$ are known at two and one loop respectively \cite{Kniehl:2002br}. $V_A , V_B = 1+ {\cal O} (\als^2)$ \cite{Brambilla:2006wp}. This lagrangian has been used to carry our calculations at fixed order in $\als$: an almost complete NNNLO (assuming $\lQ \ll m_Q\als^2$) expression for the spectrum is available 
\cite{Penin:2002zv}. Most remarkably resummations of logarithms can also be carried out using renormalization group techniques \cite{Pineda:2001ra,Pineda:2001et,Pineda:2002bv} \footnote{The importance of log resummations was first emphasized in the vNRQCD approach \cite{Luke:1999kz}. However, the correct results for the spectrum and electromagnetic decay widths were first obtained in pNRQCD \cite{Pineda:2001ra,Pineda:2001et} and later on reproduced in vNRQCD \cite{Hoang:2002yy}.}. Thus the $\eta_b$ mass and its electromagnetic decay width have been predicted at NNLL and NLL respectively \cite{Kniehl:2003ap,Penin:2004ay}.

\subsubsection{Strong Coupling Regime}

The lagrangian in this regime reads \footnote{We ignore for simplicity pseudo-Goldstone bosons.}  

\bea 
L_{\rm pNRQCD} = \int d^3 {\bf R}  \int d^3 {\bf r}  \;
S^\dagger \big( i\partial_0 - h_s({\bf r}, {\bf p}, {\bf P_R},
{\bf S}_1,  {\bf S}_2) \big) S,&&
\label{pnrqcdstrong}\nn
\eea
\bea 
h_s({\bf r}, {\bf p}, {\bf P_R}, {\bf S}_1,  {\bf S}_2) =
{{\bf p}^2\over m_Q} + {{\bf P_R}_2^2\over 4m_Q}
+ V_s({\bf r}, {\bf p}, {\bf P_R}, {\bf S}_1,  {\bf S}_2),&&
\nn
\eea
\bea
&& V_s = 
V^{(0)}_s + {V^{(1)}_s \over m_Q}
+ {V^{(2)}_s \over m_Q^2}+ \cdots ,
\label{V1ovm2}
\nn
\eea
$V_s$ cannot be calculated in perturbation theory of $\als (m_Qv)$ anymore, but can indeed be, and most of the terms have been, calculated on (quenched) lattice simulations 
\cite{Koma,Bali:1997am,Necco:2001xg,Koma:2006si,Koma:2006fw} (see \cite{Jugeau:2003df} for a large $N_c$ calculation). The fact that $\lQ \gg m_Qv^2$ can now be exploited to further factorize NRQCD matrix elements \cite{Brambilla:2001xy,Brambilla:2002nu,Brambilla:2003mu}, for instance

$$\langle \Upsilon (n)|O_8(^1S_0)|\Upsilon (n)\rangle=
C_A {|R_{n}({0})|^2 \over 2\pi}
\left(-{(C_A/2-C_f) c_F^2{\cal B}_1 \over 3 m_Q^2 }\right)   $$
were $R_{n}({0})$ is the wave function at the origin,
 ${\cal B}_1 \sim \lQ^2$ is a universal (independent of $n$) non-perturbative parameter, and 
 $c_F$ a (computable) short distance matching coefficient. This further factorization allows to put forward new model independent predictions, for instance the ratios of hadronic decay widths of P-wave states in bottomonium were predicted from charmonium data \cite{Brambilla:2001xy}, or the ratio of photon spectra in radiative decays of vector resonances \cite{GarciaiTormo:2005bs}, as we will see below.

\subsubsection{Weak or Strong (or else)?}

Since $m_Q$ , $v$ and $\lQ$ are not directly observable, given heavy quarkonium state it is not clear to which of the above regimes, if to any \footnote{States close or above the open flavor threshold are expected to belong neither the weak nor the strong coupling regimes.} , it must be assigned to. The leading non-perturbative ($\sim \lQ$) corrections to the spectrum in the weak coupling regime scale as a large power of the principal quantum number \cite{Voloshin:1978hc,Leutwyler:1980tn}, which suggests that only the $n=1$ states of bottomonium and charmonium may belong to this regime. However if one ignores this and proceeds with weak coupling calculations one finds, for instance, that renormalon based approaches at NNLO \cite{Brambilla:2001fw,Brambilla:2001qk} and the NNNLO calculation \cite{Penin:2004ay} give a reasonable description of the bottomonium spectrum up to $n=3$. It has recently been proposed that precise measurements of the photon spectra in radiative decays, as the ones carried out by CLEO \cite{Besson:2005jv}, will clarify the assignments \cite{GarciaiTormo:2005bs}. It turns out that in the strong coupling regime one can work out the following formula, which holds at NLO,
$$\frac{\displaystyle\frac{d\Gamma_n}{dz}}{\displaystyle\frac{d\Gamma_r}{dz}} 
=\frac{\langle\mathcal{O}_1(^3S_1)\rangle_n}{\langle\mathcal{O}_1(^3S_1)\rangle_r}
\left(1+\frac{C_1'\left[\phantom{}^3S_1\right](z)}{C_1\left[\phantom{}^3S_1\right](z)}\frac{1}{m_Q}\left(E_{n}-E_{r}\right)\right)
$$
$$ \frac{\langle\mathcal{O}_1(^3S_1)\rangle_n}{\langle\mathcal{O}_1(^3S_1)\rangle_r} 
=
\frac{\Gamma\left(\Upsilon (n)\to e^+e^-\right)}{\Gamma\left(\Upsilon (r)\to e^+e^-\right)}
\left[\!1\!-\!\frac{\mathrm{Im}g_{ee}\left(\phantom{}^3S_1\right)}{\mathrm{Im}f_{ee}\left(\phantom{}^3S_1\right)}\frac{E_{n}-E_{r}}{m_Q}\right]
$$
$C_1'\left[\phantom{}^3S_1\right](z)$, $C_1\left[\phantom{}^3S_1\right](z)$, $\mathrm{Im}g_{ee}\left(\phantom{}^3S_1\right)$, $\mathrm{Im}f_{ee}\left(\phantom{}^3S_1\right)$ are matching coefficients computable in perturbation theory, and $E_{n}-E_{r}$ the mass difference between the two states.
If data follow this formula it will indicate that both $n$ and $r$ are in the strong coupling regime. For the $n=1,2,3$ of bottomonia, current data disfavor $n=1$ in the strong coupling regime and is compatible with $n=2,3$ in it. 

\subsection{Soft-Collinear Effective Theory}

\begin{figure}
\resizebox{0.5\textwidth}{!}{
  \includegraphics{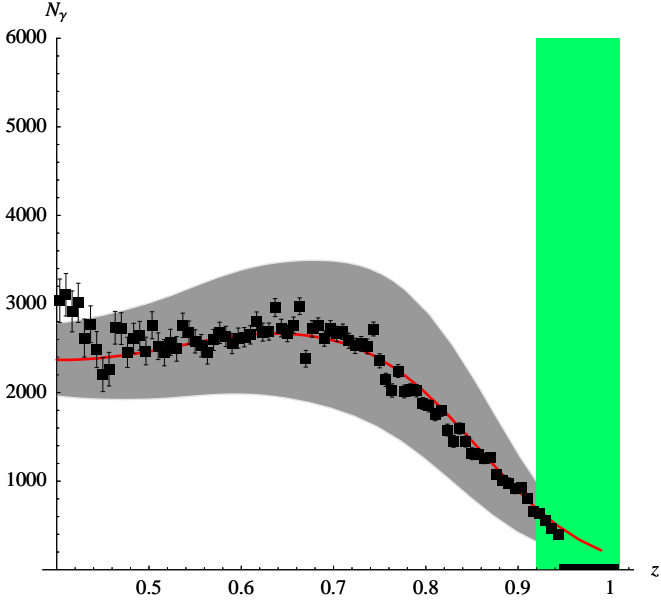}
}
\caption{$N_\gamma$ is the number of photons and $z$ the energy of the photons normalized to half the $\Upsilon (1S)$ mass. The black squares are CLEO data  \cite{Besson:2005jv} and the (red) solid curve the theoretical result of \cite{GarciaiTormo:2005ch} with theoretical errors (grey band) displayed. The (green) rectangle on the right displays the region where the theoretical results are not reliable. The plot is taken from \cite{Tormo:2006ew}.}
\label{fig:1}       
\end{figure}

For exclusive decays and for certain kinematical end-points of semi-inclusive decays, NRQCD must be supplemented with collinear degrees of freedom. This can be done in the effective theory framework of Soft-Collinear Effective Theory (SCET)\cite{Bauer:2000ew,Bauer:2000yr}. Exclusive radiative decays of heavy quarkonium in SCET have been addressed in \cite{Fleming:2004hc}, where results analogous to those of traditional light cone factorization formulas have been obtained \cite{Ma:2001tt}. Concerning semi-inclusive decays, let me focus on the photon spectrum of $\Upsilon (1S) \rightarrow \gamma X$ when the photon is very energetic. Describing experimental data in this region has been a challenge since the early days of QCD, and the best fits were obtained using a model with a finite gluon mass of the order of the GeV. \cite{Field:2001iu}. The naive inclusion of color octet contributions in the NRQCD framework seemed to further deepen the discrepancy \cite{Wolf:2000pm}. In recent years a remarkable improvement in the understanding of this end-point region has been achieved by combining  SCET and NRQCD. The following factorization formula has been proved \cite{Fleming:2002rv,Fleming:2002sr}
\begin{displaymath}
\frac{d\Gamma^e}{dz}=\sum_{\omega}H(M,\omega,\mu)\int dkS(k,\mu)\mathrm{Im}J_{\omega}(k+M(1-z),\mu)
\end{displaymath}
where $H$ is a hard matching coefficient computable in perturbation theory, $J_w$ is a jet function (also computable in perturbation theory if z is not too close to $1$) and $S$ is a (soft) shape function which depends on the bound state dynamics. $M$ is the heavy quarkonium mass and $\mu$ a factorization scale. Large (Sudakov) logs have been resummed using renormalization group equations both in the color octet \cite{Bauer:2001rh} and color singlet \cite{Fleming:2004rk} contributions. Furthermore, using pNRQCD the shape function $S$ has been calculated \cite{GarciaiTormo:2004jw}. When all this is put together an excellent description of data is achieved \cite{GarciaiTormo:2005ch}, see Fig. \ref{fig:1}. 
\section{Production}

Production of heavy quarkonium is a far more complicated issue than spectroscopy and decay. This is due to the fact that in addition to the several scales which characterize the heavy quarkonium system, further scales due to the kinematics of the production process may also appear.

\subsection{Electromagnetic threshold production}

This is the simplest and best understood production process in the weak coupling regime, which is relevant for a precise measurement of top quark mass in the future International Linear Collider. The cross-section at NNLO is known for some time \cite{Hoang:2000yr} and the log resummation at NLL is also available \cite{Pineda:2001et,Hoang:2002yy}. The effort is now in calculations at NNNLO, where partial results already exist \cite{Kniehl:1999mx,Hoang:2003ns,Kniehl:2002yv,Beneke:2005hg,Penin:2005eu,Marquard:2006qi}, and in the log resummation at NNLL, where partial results also exist \cite{Hoang:2003ns,Pineda:2006ri}. At this level of precision electroweak effect must also be taken into account \cite{Hoang:2004tg,Hoang:2006pd,Eiras:2006xm} (see \cite{Hoang:2006pc,Signer:2006vq} for recent reviews).

\subsection{Inclusive production}

A factorization formula for the inclusive production of heavy quarkonium was put forward in \cite{Bodwin:1994jh}, which was assumed to hold provided that the tranverse momentum ${\bf p_\perp}$ was larger or of the order of the heavy quark mass,  
\bea 
&&
\sigma(H)
\;=\; \sum_n {F_n(\mu) \over m_Q^{d_n-4}} \;
  {\langle 0 |} {\cal O}^H_n(\mu) {| 0 \rangle},
\nonumber
\eea
The formula contains a partonic level cross section ($F_n(\mu) / $ $m_Q^{d_n-4}$) in which the heavy quark pair may be produced in color singlet or a color octet state, and long distance production matrix elements ($ {\langle 0 |} {\cal O}^H_n(\mu) {| 0 \rangle}$) which encode the evolution of the pair to the actual physical state. The production matrix elements have the generic form
\begin{eqnarray}
{\cal O}^H_n &=&
\chi^\dagger {\cal K}_n \psi \left( \sum_X \sum_{m_J}{| H+X \rangle}
	{\langle H+X |} \right) \psi^\dagger {\cal K}_n' \chi
\nonumber
\end{eqnarray}
where the sums are over the $2J+1$ spin states of the quarkonium $H$ and
over all other final-state particles $X$. ${\cal K}_n$, ${\cal K}_n'$ are gluonic operators (including no gluon content and covariant derivatives). The production matrix elements were
assigned sizes according to the velocity scaling rules of \cite{Bodwin:1994jh}, which, as discussed above, correspond to the weak coupling regime. A great success of this formula was the explanation of charmonium production a the Tevatron \cite{Braaten:1994vv}, and it has been applied to a large number of production processes (see \cite{Kramer:2001hh,Bodwin:2003kc,Bodwin:2005ec} for reviews), including some NLO calculations in photoproduction \cite{Klasen:2004tz,Klasen:2004az}. This factorization formalism has received a closer look recently in the framework of fragmentation functions, and has been proved to be correct at NNLO in $\als (m_Q)$, provided a slight redefinition of the matrix elements is carried out \cite{Nayak:2005rw,Nayak:2005rt,Nayak:2006fm}. The interplay of the scales $m_Qv$, $m_Qv^2$ and $\lQ$ has not been discussed in detail for production. This may be important in resolving the polarization puzzle at the Tevatron \cite{Affolder:2000nn}, since the NRQCD prediction that heavy quarkonium must be produced transversely polarized in fragmentation processes \cite{Cho:1994ih}, sometimes phrased as a smoking gun for NRQCD, depends crucially on the velocity counting used. For instance, in the strong coupling regime one would not expect a sizable polarization \cite{Fleming:2000ib}. 

Near certain kinematical end-points NRQCD production processes must be supplemented with collinear degrees of freedom, in analogy to semi-inclusive decays discussed above. This issue has recently been addressed using SCET \cite{Fleming:2003gt,Hagiwara:2004pf,Lin:2004eu,Fleming:2006cd}.

\subsection{Exclusive production}

Although no detailed formalism has been worked out for exclusive production, the basic ideas of NRQCD factorization have indeed been applied to it \cite{Braaten:2002fi,Zhang:2005ch}, mostly after the surprisingly large double charmonium cross-section first measured at Belle \cite{Abe:2002rb}. The traditional light-cone factorization does not seem to have major problems to accommodate this cross-section \cite{Ma:2004qf,Bondar:2004sv,Braguta:2006py}, although some modeling of the light-cone distribution amplitudes has so far been required. Recently, a further factorization of these objects in NRQCD has been presented \cite{Ma:2006hc}, and a detailed comparison of the NRQCD and light-cone approaches has been carried out \cite{Bodwin:2006dm} (see also \cite{Braguta:2006wr}). The latter suggests that the actual cross-section may well be accommodated into the NRQCD factorization results once higher order contributions in the velocity expansion are taken into account.

\section{Conclusions}

The NRQCD formalism provides a solid QCD-based framework where heavy quarkonium spectroscopy and inclusive decays can be systematically described starting from QCD. An NRQCD factorization formalism has also been put forward for semi-inclusive decays, inclusive production and, more recently, exclusive production, which, in spite of its successes, is not so well understood theoretically. Nevertheless, remarkable progress has been done recently concerning the structure of the factorization formulas for inclusive processes and the relation to the light-cone formalism for exclusive ones. Certain kinematical end-points both in semi-inclusive decays and production require the inclusion of collinear degrees of freedom in NRQCD. Progress has also occurred here by combining NRQCD and SCET.      

\section*{Acknowledgements}
I acknowledge financial support from MEC (Spain) grant CYT FPA
2004-04582-C02-01, the CIRIT (Catalonia) grant 2005SGR00564, and the RTNs
Euridice HPRN-CT2002-00311 and Flavianet MRTN-CT-2006-035482 (EU).

%
%
%
%

\begin{thebibliography}{}
%
%
\bibitem{Brambilla:2004wf}
  N.~Brambilla {\it et al.},
  arXiv:hep-ph/0412158.

\bibitem{Caswell:1985ui}
  W.~E.~Caswell and G.~P.~Lepage,
  Phys.\ Lett.\ B {\bf 167} (1986) 437.

\bibitem{Bodwin:1994jh}
  G.~T.~Bodwin, E.~Braaten and G.~P.~Lepage,
  Phys.\ Rev.\ D {\bf 51} (1995) 1125
  [Erratum-ibid.\ D {\bf 55} (1997) 5853]
  [arXiv:hep-ph/9407339].


\bibitem{Brambilla:2004jw}
  N.~Brambilla, A.~Pineda, J.~Soto and A.~Vairo,
  Rev.\ Mod.\ Phys.\  {\bf 77} (2005) 1423
  [arXiv:hep-ph/0410047].


\bibitem{Brambilla:2005yk}
  N.~Brambilla, A.~Vairo and T.~Rosch,
  Phys.\ Rev.\ D {\bf 72} (2005) 034021
  [arXiv:hep-ph/0506065].

\bibitem{Fleming:2005pd}
  S.~Fleming and T.~Mehen,
  Phys.\ Rev.\ D {\bf 73} (2006) 034502
  [arXiv:hep-ph/0509313].

\bibitem{Pineda:1997bj}
  A.~Pineda and J.~Soto,
  Nucl.\ Phys.\ Proc.\ Suppl.\  {\bf 64} (1998) 428
  [arXiv:hep-ph/9707481].

\bibitem{Brambilla:1999xf}
  N.~Brambilla, A.~Pineda, J.~Soto and A.~Vairo,
  Nucl.\ Phys.\ B {\bf 566} (2000) 275
  [arXiv:hep-ph/9907240].

\bibitem{Neubert:1993mb}
  M.~Neubert,
  Phys.\ Rept.\  {\bf 245} (1994) 259
  [arXiv:hep-ph/9306320].


\bibitem{Vairo:2006nq}
  A.~Vairo, these proceedings,
  arXiv:hep-ph/0610251.

\bibitem{Gray:2005ur}
  A.~Gray, I.~Allison, C.~T.~H.~Davies, E.~Dalgic, G.~P.~Lepage, J.~Shigemitsu and M.~Wingate,
  Phys.\ Rev.\ D {\bf 72} (2005) 094507
  [arXiv:hep-lat/0507013].

\bibitem{Bodwin:1992ye}
  G.~T.~Bodwin, E.~Braaten and G.~P.~Lepage,
  Phys.\ Rev.\ D {\bf 46} (1992) 1914
  [arXiv:hep-lat/9205006].

\bibitem{Bodwin:2001mk}
  G.~T.~Bodwin, D.~K.~Sinclair and S.~Kim,
  Phys.\ Rev.\ D {\bf 65} (2002) 054504
  [arXiv:hep-lat/0107011].

\bibitem{Petrelli:1997ge}
  A.~Petrelli, M.~Cacciari, M.~Greco, F.~Maltoni and M.~L.~Mangano,
  Nucl.\ Phys.\ B {\bf 514} (1998) 245
  [arXiv:hep-ph/9707223].

\bibitem{Brambilla:2006ph}
  N.~Brambilla, E.~Mereghetti and A.~Vairo,
  JHEP {\bf 0608} (2006) 039
  [arXiv:hep-ph/0604190].

\bibitem{Bodwin:2002hg}
  G.~T.~Bodwin and A.~Petrelli,
  Phys.\ Rev.\ D {\bf 66} (2002) 094011
  [arXiv:hep-ph/0205210].


\bibitem{Ma:2002ev}
  J.~P.~Ma and Q.~Wang,
  Phys.\ Lett.\ B {\bf 537}, 233 (2002)
  [arXiv:hep-ph/0203082].

\bibitem{Luke:1999kz}
  M.~E.~Luke, A.~V.~Manohar and I.~Z.~Rothstein,
  Phys.\ Rev.\ D {\bf 61} (2000) 074025
  [arXiv:hep-ph/9910209].

\bibitem{Griesshammer:1997wz}
  H.~W.~Griesshammer,
  Phys.\ Rev.\ D {\bf 58} (1998) 094027
  [arXiv:hep-ph/9712467].

\bibitem{Manohar:2006nz}
  A.~V.~Manohar and I.~W.~Stewart,
  arXiv:hep-ph/0605001.

\bibitem{Hoang:2002yy}
  A.~H.~Hoang and I.~W.~Stewart,
  Phys.\ Rev.\ D {\bf 67} (2003) 114020
  [arXiv:hep-ph/0209340].

\bibitem{Manohar} A. Manohar, these proceedings.

\bibitem{Peter:1997me}
  M.~Peter,
  Nucl.\ Phys.\ B {\bf 501} (1997) 471
  [arXiv:hep-ph/9702245].

\bibitem{Schroder:1998vy}
  Y.~Schroder,
  Phys.\ Lett.\ B {\bf 447} (1999) 321
  [arXiv:hep-ph/9812205].

\bibitem{Kniehl:2004rk}
  B.~A.~Kniehl, A.~A.~Penin, Y.~Schroder, V.~A.~Smirnov and M.~Steinhauser,
  Phys.\ Lett.\ B {\bf 607} (2005) 96
  [arXiv:hep-ph/0412083].

\bibitem{Brambilla:1999qa}
  N.~Brambilla, A.~Pineda, J.~Soto and A.~Vairo,
  Phys.\ Rev.\ D {\bf 60} (1999) 091502
  [arXiv:hep-ph/9903355].

\bibitem{Brambilla:2006wp}
  N.~Brambilla, X.~Garcia~i~Tormo, J.~Soto and A.~Vairo,
  arXiv:hep-ph/0610143.

\bibitem{Pineda:2002se}
  A.~Pineda,
  J.\ Phys.\ G {\bf 29} (2003) 371
  [arXiv:hep-ph/0208031].

\bibitem{Bali:2003jq}
  G.~S.~Bali and A.~Pineda,
  Phys.\ Rev.\ D {\bf 69} (2004) 094001
  [arXiv:hep-ph/0310130].

\bibitem{Kniehl:2002br}
  B.~A.~Kniehl, A.~A.~Penin, V.~A.~Smirnov and M.~Steinhauser,
  Nucl.\ Phys.\ B {\bf 635} (2002) 357
  [arXiv:hep-ph/0203166].

\bibitem{Penin:2002zv}
  A.~A.~Penin and M.~Steinhauser,
  Phys.\ Lett.\ B {\bf 538} (2002) 335
  [arXiv:hep-ph/0204290].

\bibitem{Pineda:2001ra}
  A.~Pineda,
  Phys.\ Rev.\ D {\bf 65} (2002) 074007
  [arXiv:hep-ph/0109117].

\bibitem{Pineda:2001et}
  A.~Pineda,
  Phys.\ Rev.\ D {\bf 66} (2002) 054022
  [arXiv:hep-ph/0110216].

\bibitem{Pineda:2002bv}
  A.~Pineda,
  Phys.\ Rev.\ A {\bf 66} (2002) 062108
  [arXiv:hep-ph/0204213].

\bibitem{Kniehl:2003ap}
  B.~A.~Kniehl, A.~A.~Penin, A.~Pineda, V.~A.~Smirnov and M.~Steinhauser,
  Phys.\ Rev.\ Lett.\  {\bf 92} (2004) 242001
  [arXiv:hep-ph/0312086].

\bibitem{Penin:2004ay}
  A.~A.~Penin, A.~Pineda, V.~A.~Smirnov and M.~Steinhauser,
  Nucl.\ Phys.\ B {\bf 699} (2004) 183
  [arXiv:hep-ph/0406175].



\bibitem{Koma} Y. Koma, these proceedings.

\bibitem{Bali:1997am}
  G.~S.~Bali, K.~Schilling and A.~Wachter,
  Phys.\ Rev.\ D {\bf 56} (1997) 2566
  [arXiv:hep-lat/9703019].

\bibitem{Necco:2001xg}
  S.~Necco and R.~Sommer,
  Nucl.\ Phys.\ B {\bf 622} (2002) 328
  [arXiv:hep-lat/0108008].

\bibitem{Koma:2006si}
  Y.~Koma, M.~Koma and H.~Wittig,
  Phys.\ Rev.\ Lett.\  {\bf 97}, 122003 (2006)
  [arXiv:hep-lat/0607009].

\bibitem{Koma:2006fw}
  Y.~Koma and M.~Koma,
  arXiv:hep-lat/0609078.

\bibitem{Jugeau:2003df}
  F.~Jugeau and H.~Sazdjian,
  Nucl.\ Phys.\ B {\bf 670} (2003) 221.

\bibitem{Brambilla:2001xy}
  N.~Brambilla, D.~Eiras, A.~Pineda, J.~Soto and A.~Vairo,
  Phys.\ Rev.\ Lett.\  {\bf 88} (2002) 012003
  [arXiv:hep-ph/0109130].

\bibitem{Brambilla:2002nu}
  N.~Brambilla, D.~Eiras, A.~Pineda, J.~Soto and A.~Vairo,
  Phys.\ Rev.\ D {\bf 67} (2003) 034018
  [arXiv:hep-ph/0208019].

\bibitem{Brambilla:2003mu}
  N.~Brambilla, A.~Pineda, J.~Soto and A.~Vairo,
  Phys.\ Lett.\ B {\bf 580} (2004) 60
  [arXiv:hep-ph/0307159].

\bibitem{GarciaiTormo:2005bs}
  X.~Garcia i Tormo and J.~Soto,
  Phys.\ Rev.\ Lett.\  {\bf 96} (2006) 111801
  [arXiv:hep-ph/0511167].

\bibitem{Voloshin:1978hc}
  M.~B.~Voloshin,
  Nucl.\ Phys.\ B {\bf 154} (1979) 365.

\bibitem{Leutwyler:1980tn}
  H.~Leutwyler,
  Phys.\ Lett.\ B {\bf 98} (1981) 447.

\bibitem{Brambilla:2001fw}
  N.~Brambilla, Y.~Sumino and A.~Vairo,
  Phys.\ Lett.\ B {\bf 513} (2001) 381
  [arXiv:hep-ph/0101305].

\bibitem{Brambilla:2001qk}
  N.~Brambilla, Y.~Sumino and A.~Vairo,
  Phys.\ Rev.\ D {\bf 65} (2002) 034001
  [arXiv:hep-ph/0108084].

\bibitem{Besson:2005jv}
  D.~Besson {\it et al.}  [CLEO Collaboration],
  Phys.\ Rev.\ D {\bf 74} (2006) 012003
  [arXiv:hep-ex/0512061].


\bibitem{Bauer:2000ew}
  C.~W.~Bauer, S.~Fleming and M.~E.~Luke,
  Phys.\ Rev.\ D {\bf 63}, 014006 (2001)
  [arXiv:hep-ph/0005275].

\bibitem{Bauer:2000yr}
  C.~W.~Bauer, S.~Fleming, D.~Pirjol and I.~W.~Stewart,
  Phys.\ Rev.\ D {\bf 63}, 114020 (2001)
  [arXiv:hep-ph/0011336].

\bibitem{Fleming:2004hc}
  S.~Fleming, C.~Lee and A.~K.~Leibovich,
  Phys.\ Rev.\ D {\bf 71}, 074002 (2005)
  [arXiv:hep-ph/0411180].

\bibitem{Ma:2001tt}
  J.~P.~Ma,
  Nucl.\ Phys.\ B {\bf 605}, 625 (2001)
  [Erratum-ibid.\ B {\bf 611}, 523 (2001)]
  [arXiv:hep-ph/0103237].

\bibitem{Field:2001iu}
  J.~H.~Field,
  Phys.\ Rev.\ D {\bf 66}, 013013 (2002).

\bibitem{Wolf:2000pm}
  S.~Wolf,
  Phys.\ Rev.\ D {\bf 63}, 074020 (2001).

\bibitem{Fleming:2002rv}
  S.~Fleming and A.~K.~Leibovich,
  Phys.\ Rev.\ Lett.\  {\bf 90}, 032001 (2003)
  [arXiv:hep-ph/0211303].

\bibitem{Fleming:2002sr}
  S.~Fleming and A.~K.~Leibovich,
  Phys.\ Rev.\ D {\bf 67}, 074035 (2003)
  [arXiv:hep-ph/0212094].

\bibitem{Bauer:2001rh}
  C.~W.~Bauer, C.~W.~Chiang, S.~Fleming, A.~K.~Leibovich and I.~Low,
  Phys.\ Rev.\ D {\bf 64}, 114014 (2001).

\bibitem{Fleming:2004rk}
  S.~Fleming and A.~K.~Leibovich,
  Phys.\ Rev.\ D {\bf 70}, 094016 (2004)
  [arXiv:hep-ph/0407259].

\bibitem{GarciaiTormo:2004jw}
  X.~Garcia i Tormo and J.~Soto,
  Phys.\ Rev.\ D {\bf 69}, 114006 (2004)
  [arXiv:hep-ph/0401233].

\bibitem{GarciaiTormo:2005ch}
  X.~Garcia i Tormo and J.~Soto,
  Phys.\ Rev.\ D {\bf 72}, 054014 (2005)
  [arXiv:hep-ph/0507107].

\bibitem{Tormo:2006ew}
  X.~Garcia~i~Tormo,
  arXiv:hep-ph/0610145.

\bibitem{Hoang:2000yr}
  A.~H.~Hoang {\it et al.},
  Eur.\ Phys.\ J.\ directC {\bf 2}, 1 (2000).


\bibitem{Kniehl:1999mx}
  B.~A.~Kniehl and A.~A.~Penin,
  Nucl.\ Phys.\ B {\bf 577}, 197 (2000)
  [arXiv:hep-ph/9911414].

\bibitem{Kniehl:2002yv}
  B.~A.~Kniehl, A.~A.~Penin, M.~Steinhauser and V.~A.~Smirnov,
  Phys.\ Rev.\ Lett.\  {\bf 90}, 212001 (2003).

\bibitem{Hoang:2003ns}
  A.~H.~Hoang,
  Phys.\ Rev.\ D {\bf 69}, 034009 (2004).

\bibitem{Beneke:2005hg}
  M.~Beneke, Y.~Kiyo and K.~Schuller,
  Nucl.\ Phys.\ B {\bf 714}, 67 (2005)
  [arXiv:hep-ph/0501289].

\bibitem{Penin:2005eu}
  A.~A.~Penin, V.~A.~Smirnov and M.~Steinhauser,
  Nucl.\ Phys.\ B {\bf 716}, 303 (2005)
  [arXiv:hep-ph/0501042].

\bibitem{Marquard:2006qi}
  P.~Marquard, J.~H.~Piclum, D.~Seidel and M.~Steinhauser,
  arXiv:hep-ph/0607168.

\bibitem{Pineda:2006ri}
  A.~Pineda and A.~Signer,
  arXiv:hep-ph/0607239.

\bibitem{Hoang:2004tg}
  A.~H.~Hoang and C.~J.~Reisser,
  Phys.\ Rev.\ D {\bf 71}, 074022 (2005)
  [arXiv:hep-ph/0412258].

\bibitem{Hoang:2006pd}
  A.~H.~Hoang and C.~J.~Reisser,
  Phys.\ Rev.\ D {\bf 74}, 034002 (2006)
  [arXiv:hep-ph/0604104].

\bibitem{Eiras:2006xm}
  D.~Eiras and M.~Steinhauser,
  Nucl.\ Phys.\ B {\bf 757}, 197 (2006)
  [arXiv:hep-ph/0605227].

\bibitem{Hoang:2006pc}
  A.~H.~Hoang,
  PoS {\bf TOP2006}, 032 (2006)
  [arXiv:hep-ph/0604185].

\bibitem{Signer:2006vq}
  A.~Signer,
  PoS {\bf TOP2006}, 033 (2006)
  [arXiv:hep-ph/0604032].

\bibitem{Braaten:1994vv}
  E.~Braaten and S.~Fleming,
  Phys.\ Rev.\ Lett.\  {\bf 74}, 3327 (1995)
  [arXiv:hep-ph/9411365].

\bibitem{Kramer:2001hh}
  M.~Kramer,
  Prog.\ Part.\ Nucl.\ Phys.\  {\bf 47}, 141 (2001).

\bibitem{Bodwin:2003kc}
  G.~T.~Bodwin, J.~Lee and R.~Vogt,
  arXiv:hep-ph/0305034.

\bibitem{Bodwin:2005ec}
  G.~T.~Bodwin,
  Int.\ J.\ Mod.\ Phys.\ A {\bf 21}, 785 (2006).

\bibitem{Klasen:2004tz}
  M.~Klasen, B.~A.~Kniehl, L.~N.~Mihaila and M.~Steinhauser,
  Nucl.\ Phys.\ B {\bf 713}, 487 (2005)
  [arXiv:hep-ph/0407014].

\bibitem{Klasen:2004az}
  M.~Klasen, B.~A.~Kniehl, L.~N.~Mihaila and M.~Steinhauser,
  Phys.\ Rev.\ D {\bf 71}, 014016 (2005)
  [arXiv:hep-ph/0408280].

\bibitem{Nayak:2005rw}
  G.~C.~Nayak, J.~W.~Qiu and G.~Sterman,
  Phys.\ Lett.\ B {\bf 613}, 45 (2005)
  [arXiv:hep-ph/0501235].

\bibitem{Nayak:2005rt}
  G.~C.~Nayak, J.~W.~Qiu and G.~Sterman,
  Phys.\ Rev.\ D {\bf 72}, 114012 (2005)
  [arXiv:hep-ph/0509021].

\bibitem{Nayak:2006fm}
  G.~C.~Nayak, J.~W.~Qiu and G.~Sterman,
  Phys.\ Rev.\ D {\bf 74}, 074007 (2006)
  [arXiv:hep-ph/0608066].

\bibitem{Affolder:2000nn}
  A.~A.~Affolder {\it et al.}  [CDF Collaboration],
  Phys.\ Rev.\ Lett.\  {\bf 85} (2000) 2886
  [arXiv:hep-ex/0004027].

\bibitem{Cho:1994ih}
  P.~L.~Cho and M.~B.~Wise,
  Phys.\ Lett.\ B {\bf 346}, 129 (1995).

\bibitem{Fleming:2000ib}
  S.~Fleming, I.~Z.~Rothstein and A.~K.~Leibovich,
  Phys.\ Rev.\ D {\bf 64}, 036002 (2001)
  [arXiv:hep-ph/0012062].

\bibitem{Fleming:2003gt}
  S.~Fleming, A.~K.~Leibovich and T.~Mehen,
  Phys.\ Rev.\ D {\bf 68}, 094011 (2003)
  [arXiv:hep-ph/0306139].

\bibitem{Hagiwara:2004pf}
  K.~Hagiwara, E.~Kou, Z.~H.~Lin, C.~F.~Qiao and G.~H.~Zhu,
  Phys.\ Rev.\ D {\bf 70}, 034013 (2004)
  [arXiv:hep-ph/0401246].

\bibitem{Lin:2004eu}
  Z.~H.~Lin and G.~h.~Zhu,
  Phys.\ Lett.\ B {\bf 597}, 382 (2004).


\bibitem{Fleming:2006cd}
  S.~Fleming, A.~K.~Leibovich and T.~Mehen,
  arXiv:hep-ph/0607121.

\bibitem{Braaten:2002fi}
  E.~Braaten and J.~Lee,
  Phys.\ Rev.\ D {\bf 67}, 054007 (2003)
  [Erratum-ibid.\ D {\bf 72}, 099901 (2005)]
  [arXiv:hep-ph/0211085].

\bibitem{Zhang:2005ch}
  Y.~J.~Zhang, Y.~j.~Gao and K.~T.~Chao,
  Phys.\ Rev.\ Lett.\  {\bf 96} (2006) 092001
  [arXiv:hep-ph/0506076].

\bibitem{Abe:2002rb}
  K.~Abe {\it et al.}  [Belle Collaboration],
  Phys.\ Rev.\ Lett.\  {\bf 89}, 142001 (2002)
  [arXiv:hep-ex/0205104].

\bibitem{Ma:2004qf}
  J.~P.~Ma and Z.~G.~Si,
  Phys.\ Rev.\ D {\bf 70}, 074007 (2004).

\bibitem{Bondar:2004sv}
  A.~E.~Bondar and V.~L.~Chernyak,
  Phys.\ Lett.\ B {\bf 612}, 215 (2005)
  [arXiv:hep-ph/0412335].

\bibitem{Braguta:2006py}
  V.~V.~Braguta, A.~K.~Likhoded and A.~V.~Luchinsky,
  Phys.\ Rev.\ D {\bf 74} (2006) 094004
  [arXiv:hep-ph/0602232].

\bibitem{Ma:2006hc}
  J.~P.~Ma and Z.~G.~Si,
  arXiv:hep-ph/0608221.

\bibitem{Bodwin:2006dm}
  G.~T.~Bodwin, D.~Kang and J.~Lee,
  arXiv:hep-ph/0603185.

\bibitem{Braguta:2006wr}
  V.~V.~Braguta, A.~K.~Likhoded and A.~V.~Luchinsky,
  arXiv:hep-ph/0611021.



\end{thebibliography}
%

\end{document}